\documentclass[aps,prl,twocolumn,superscriptaddress,showpacs]{revtex4}
\usepackage{CJK}
\usepackage[Symbol]{upgreek}
\usepackage{bm}
\usepackage{amsmath}
\usepackage{graphicx}
\usepackage{mathptmx} 
\usepackage{endnotes}
\usepackage{color}
\usepackage{float}
\usepackage[FIGTOPCAP, hang, tight ]{subfigure}
\begin{document}
\begin{CJK*}{GB}{song}


\title{Constructing a Large Variety of Dirac-Cone Materials in the Bi${}_{1-x}$Sb${}_{x}$ Thin Film System }

\author{Shuang Tang}
\affiliation{Department of Materials Science and Engineering, Massachusetts Institute of Technology, Cambridge, MA
02139-4037, USA} 
\author{M. S. Dresselhaus}
\email[]{millie@mgm.mit.edu}
\affiliation{Department of Electrical Engineering and Computer
Science, Massachusetts Institute of Technology, Cambridge, MA
02139-4037, USA} \affiliation{Department of Physics, Massachusetts
Institute of Technology, Cambridge, MA 02139-4037, USA}

\date{\today}

\begin{abstract}
We theoretically predict that a large variety of Dirac-cone materials can be constructed in Bi${}_{1-x}$Sb${}_{x}$ thin films, and we here show how to construct single-, bi- and tri- Dirac-cone materials with various amounts of wave vector anisotropy.  These different types of Dirac cones can be of special interest to electronic devices design, quantum electrodynamics and other fields.
\end{abstract}
\maketitle
\end{CJK*}
\indent Dirac cone materials have recently attracted considerable attention. 
In an electronic band structure, if the dispersion relation $E(\mathbf{k})$ can be described by a linear function as $E=\mathbf{v}\cdot \mathbf{k}$, where $\mathbf{v}$ is the velocity, $\mathbf{k}$ is the lattice momentum, and $\hbar=1$, the point where $E\to 0$ is called a  Dirac point. A Dirac cone is a two-dimensional (2D) Dirac point. Dirac cone materials are interesting in electronic device design, quantum electrodynamics and desktop relativistic particle experiments etc. A single-, bi- or tri-Dirac cone system has one, two or three different Dirac cones degenerate in  $E(\mathbf{k})$ in the first Brillouin zone. Graphene has two degenerate isotropic Dirac cones at points $\mathit{K}$ and $\mathit{K'}$ in its first Brillouin zone, which is therefore considered as a bi-Dirac-cone system. Many novel phenomena are observed in this system \cite{1}, such as the room temperature anomalous integer quantum Hall effect \cite{2}, the Klein paradox \cite{3}, which means that fermions around a Dirac cone can transmit through a classically forbidden region with a probability of 1. Dirac fermions can be immune to localization effects 
and can propagate without scattering over large distances on the order of micrometers \cite{4}.

\begin{figure}
        \includegraphics[width=0.3\textwidth]{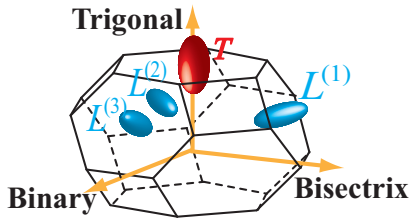}
        \caption{The 3-fold degenerate \textit{L}-point electron pockets and the \textit{T}-point hole pocket in the first Brillouin zone of bulk Bi${}_{1-x}$Sb${}_{x}$.}
\end{figure}
\begin{figure}
        \includegraphics[width=0.4 \textwidth]{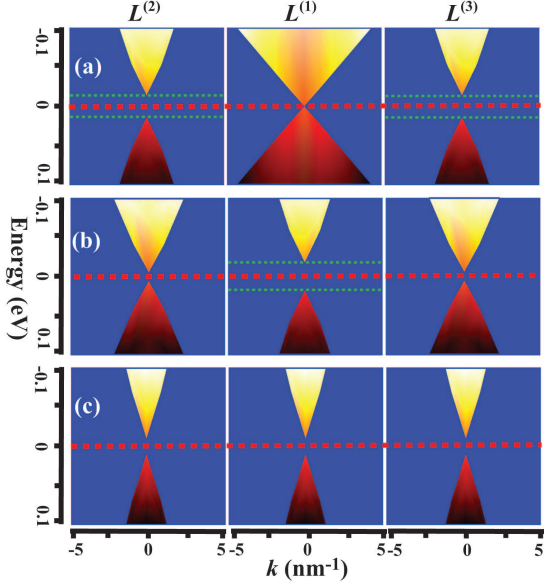}
        \caption{ An illustration of (a) single-, (b) bi- and (c) tri-Dirac-cone  Bi${}_{1-x}$Sb${}_{x}$ thin films grown along the (a) bisectrix, (b) binary and (c) trigonal axes, respectively. For the cross-sectional view of each cone, $\mathbf{k}$ is chosen such that $\nabla_{\mathbf{k}} E(\mathbf{k})$ has its minimum along that direction of $\mathbf{k}$. The illustration is based on the example of Bi${}_{1-x}$Sb${}_{x}$ thin films with $l_z=100$ nm, $x=0.04$, $P=1$ atm and $T\le77$ K, under which the \textit{L} points of bulk Bi${}_{1-x}$Sb${}_{x}$ have a zero-gap. The scenario is similar for other conditions. In (a), a single-Dirac-cone 
is formed at the $L^{(1)} $ point, while the $L^{(2)}$- 
and $L^{(3)} $- point band gaps are opened up. In (b), 
two degenerate quasi-Dirac cones are formed at the $L^{(2)} $ and $L^{(3)} $ points, while the $L^{(1)}$-point band gap 
is much larger, which leads to a bi-quasi-Dirac-cone material. 
The band gap at the $L^{(2)}$ and $L^{(3)}$ points can be less than 1 meV if a sample of $l_z=200$ nm is chosen, which leads to exact Dirac cones. In (c), the $L^{(1)} $-, $L^{(2)}$- and $L^{(3)} $- point band gaps are all the 
same, and the three quasi-Dirac cones are degenerate in energy.}
\end{figure}

\indent In this Letter, we show how to obtain single-, bi- and tri-Dirac-cone Bi${}_{1-x}$Sb${}_{x}$ thin films, and how to construct Dirac cones with different anisotropies.  We also point out the possibility of constructing semi-Dirac cones in Bi${}_{1-x}$Sb${}_{x}$ thin films.

\indent Bi${}_{1-x}$Sb${}_{x}$ has many special properties that are interesting from the point of view of anisotropic Dirac cones. We recall that bulk Bi${}_{1-x}$Sb${}_{x}$ is a crystalline alloy with a rhombohedral structure, which displays remarkable anisotropy. The first Brillouin zone of bulk Bi${}_{1-x}$Sb${}_{x}$ has one \textit{T} point and three degenerate \textit{L} points, $L^{(1)} $, $L^{(2)} $ and $L^{
(3)} $, as illustrated in Fig. 1. The bottom of the conduction band is located at the \textit{L} points, while the top of the valence band can be located either at the \textit{T} point or at the \textit{L} points, depending on the Sb composition \textit{x} when $0\le x\le 0.10$. In bulk Bi${}_{1-x}$Sb${}_{x}$, the band structure varies as 
a function of Sb compostion \textit{x}, temperature \textit{T}, pressure \textit{P} and stress $\tau$ \cite{7}. The conduction band is very close to the valence band at the \textit{L} points, so that these bands are non-parabolically dispersed as \cite{6}
\begin{equation} \label{GrindEQ__1_} 
E(\mathbf{k})=\pm ((\mathbf{v}\cdot \mathbf{k})^{2} +E_{g}^{2} )^{\frac{1}{2} }  
\end{equation} 
due to their strong interband coupling. When the \textit{L}-point band gap $E_{g}$ is small, the dispersion relation $E(\mathbf{k})$ becomes linear and Dirac points are formed as $E(\mathbf{k})\to \pm \mathbf{v}\cdot \mathbf{k}$. The \textit{L}-point band gap $E_{g}$ can approach 0 under some conditions, e.g. when $P=1$ atm, $E_{g}\to 0$ at $x \approx 0.04$ and $T \le$ 77 K \cite{8}, or at $x \approx 0.02$ and $T \le$ 300 K \cite{9}.  For simplification, this Letter will focus on the low temperature range ($T\le 77$ K) where the band structure of Bi${}_{1-x}$Sb${}_{x}$ does not change much with temperature. 
\begin{figure}
        \includegraphics[width=0.4\textwidth]{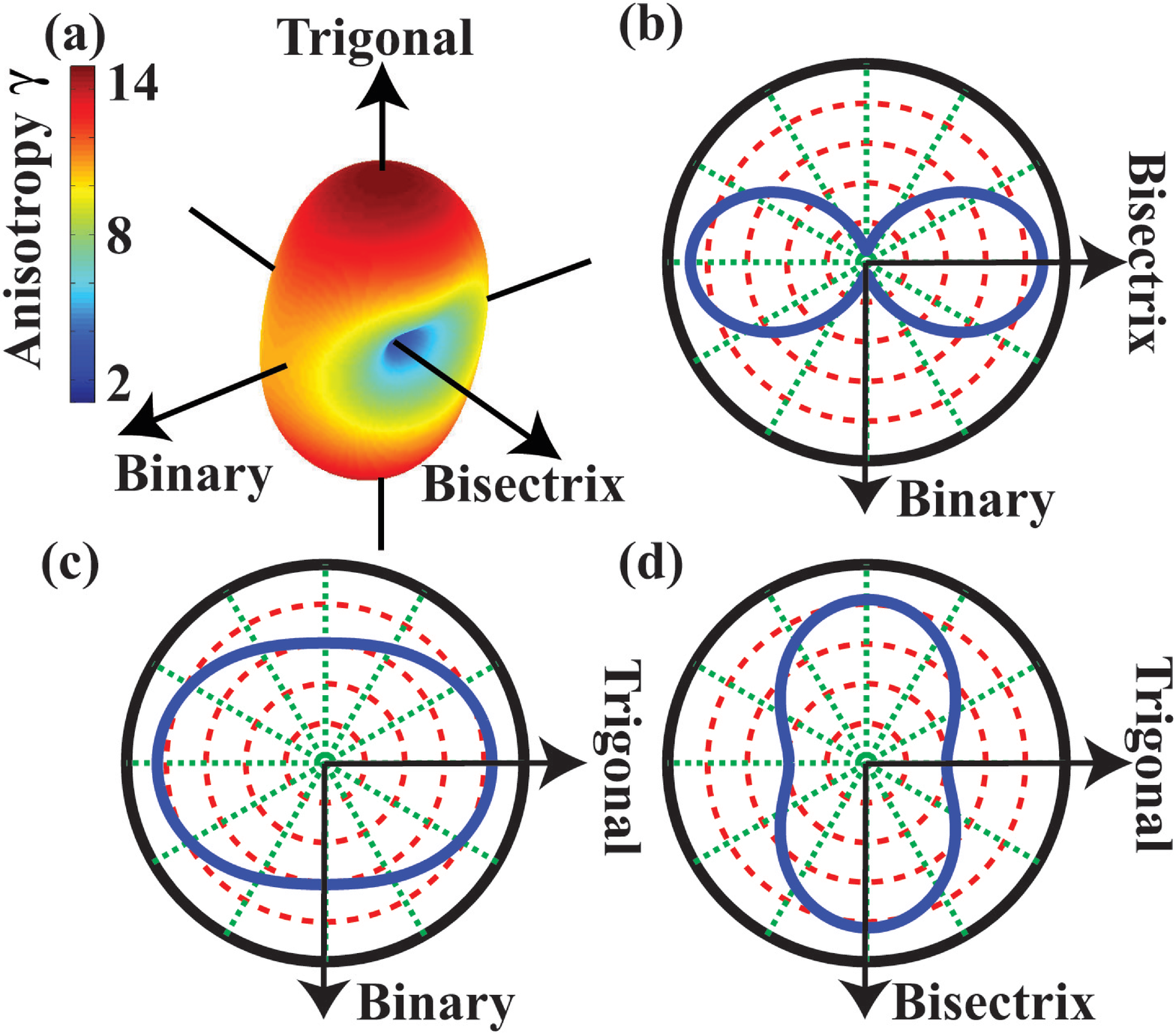}
        \caption{The anisotropy of the $L^{(1)}$-point Dirac cone 
vs. film growth orientation. (a) The anisotropy coefficient $\gamma$ vs. film growth orientation. The value of $\gamma$ for a specific film growth orientation is shown by the radius and color. $\gamma$ can be as large as $\tiny{\sim}{14}$ for films grown along the trigonal axis, and as small as $\tiny{\sim}{2}$ for films grown along the bisectrix axis. $\mathbf{v}$ is shown for the $L^{(1)} $-point Dirac fermions vs. transport direction for Bi${}_{1-x}$Sb${}_{x}$ thin films 
grown along the (b) trigonal, (c) bisectrix and (d) binary axes. (b)-(c) are drawn based on an example sample with $l_z=300$ nm and $x=0.04$.}
\end{figure}

\indent For Bi${}_{1-x}$Sb${}_{x}$ thin films, the 2D band structure varies also as a function of film thickness and film growth orientation, which provides considerable flexibility compared to bulk Bi${}_{1-x}$Sb${}_{x}$. Furthermore, the quantum confinement effect in the thin film system is potentially interesting, where its anisotropic properties imply potential application possibilities.
\begin{figure}
        \includegraphics[width=0.5\textwidth]{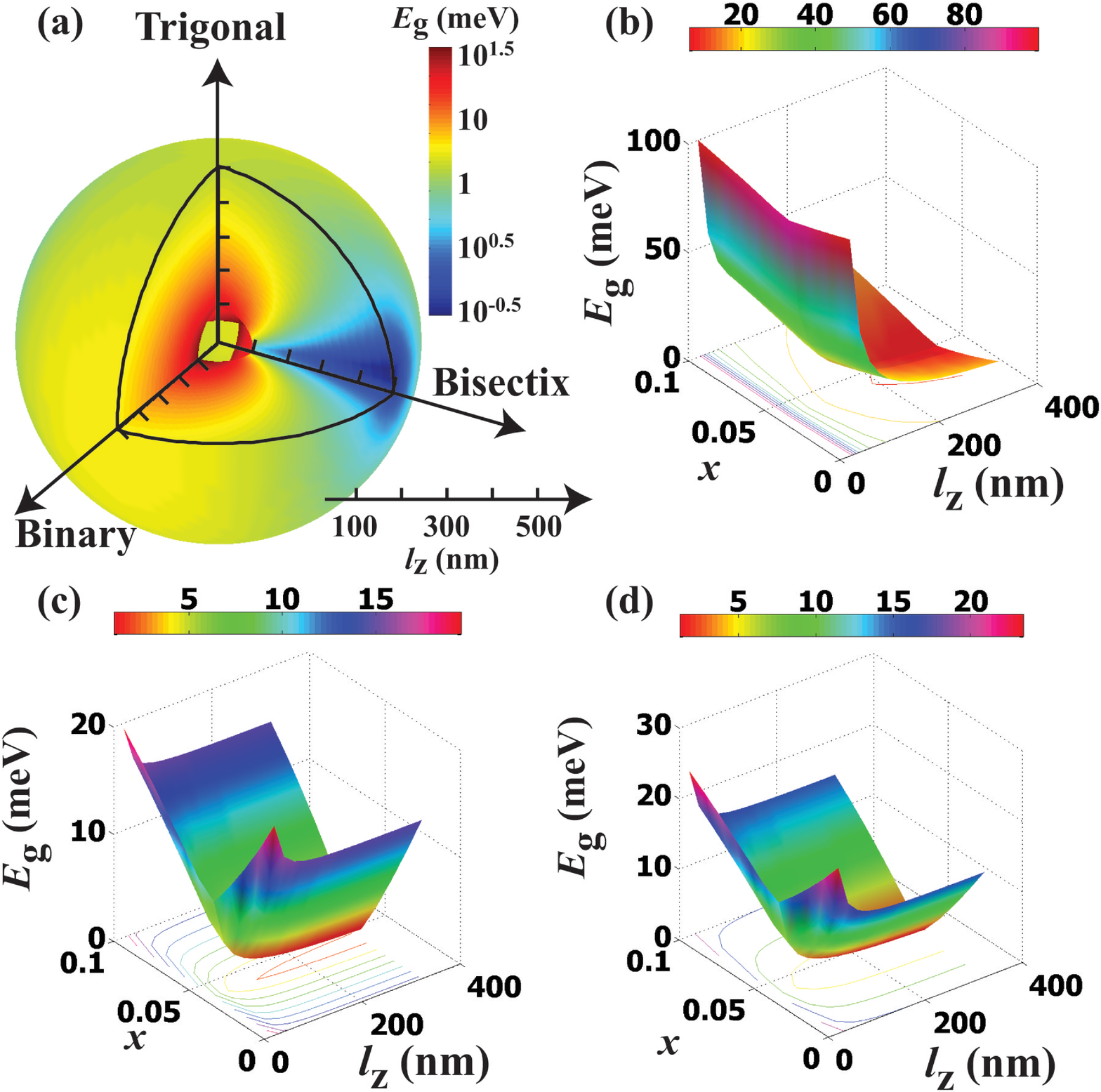}
        \caption{Scheme for the $L^{(1)}$-point band gap vs. film growth orientation, film thickness and Sb composition. (a)Illustration of the $L^{(1)}$-point band gap vs. film growth orientation and film thickness. The radius, direction and color represent the film thickness, film growth orientation and $L^{(1)} $-point band gap, respectively. The illustration takes $x=0.04$ as an example. For other Sb compositions ($0\le x\le 0.10$), the film thickness and film growth orientation dependence for the $L^{(1)}$-point band gap should be similar, which is here illustrated for thin films grown along the (b) trigonal, (c) bisectrix and (d) binary axes.}
\end{figure}

\indent The energy spectrum near an \textit{L}-point Dirac cone in a Bi${}_{1-x}$Sb${}_{x}$ thin film is calculated based on the iterative-two-dimensional-two-band model described below. Here the general two band 
model for two strongly coupled bands obeys the relation \cite{11}
\begin{equation} \label{GrindEQ__2_} 
\mathbf{p}\cdot \bm\upalpha \cdot \mathbf{p}=E(\mathbf{k})(1+\frac{E(\mathbf{k})}{E_{g} } ),                  
\end{equation} 
where $\mathbf{p}$ is the carrier momentum vector and $\bm\upalpha$ is the inverse-mass tensor. The two coupled key parameters $\bm\upalpha$ and $E_{g}$ are calculated in an iterative way 
in our model as
\begin{equation} \label{GrindEQ__3_} 
\bm\upalpha^{[n]} =\frac{E_{_{g} }^{[n-1]} }{E_{_{g} }^{[n]} } \cdot \bm\upalpha ^{[n-1]} 
+\frac{1}{m_{0} } \cdot (1-\frac{E_{_{g} }^{[n-1]} }{E_{_{g} }^{[n]} } )\cdot\mathbf{I} 
\end{equation} 
and
\begin{equation} 
\label{GrindEQ__4_} 
E_{g}^{[n+1]} =E_{g}^{[n]} +2\cdot \frac{\pi^{2} \alpha _{33}^{[n]} }{2\cdot l_{z}^{2} 
} , 
\end{equation} 
where $m_{0}$ is the free electron mass, $\mathbf{I}$ is the identity matrix, $l_{z}$ is the film thickness and  \textit{n} denotes the \textit{n}th step in the iteration. The procedure is repeated until $\bm\upalpha ^{[n]} $ and $E_{g}^{[n]} $ become self-consistent, and then we get accurate solutions for $\bm\upalpha ^{film} (Bi_{1-x} Sb_{x} )=\bm\upalpha ^{[n]} $ and $E_{g}^{film} 
(Bi_{1-x} Sb_{x} )=E_{g}^{[n]}$ for thin film Bi${}_{1-x}$Sb${}_{x}$. Because of the approximations that are valid for Sb composition $0 \le x \le 0.10$,
\[\frac{1}{m_{0} } \cdot (1-\frac{E_{_{g} }^{[n-1]} }{E_{_{g} }^{[n]} } )\cdot \mathbf{I}\ll \frac{E_{_{g} 
}^{[n-1]} }{E_{_{g} }^{[n]} } \cdot \bm\upalpha ^{[n-1]}\] 
and 
\[\left|E_{g}^{[n+1]} -E_{g}^{[n]} 
\right|\ll 2\cdot \frac{\pi^{2} \upalpha _{33}^{[n]} }{2\cdot l_{z}^{2} },\] 
Eqs. \eqref{GrindEQ__3_} and \eqref{GrindEQ__4_} can be further simplified, which converge to the analytical 
solution as
\begin{equation} \label{GrindEQ__5_} 
\bm\upalpha ^{film} (Bi_{1-x} Sb_{x} )=\frac{\bm\upalpha^{bulk} (Bi)}{E_{g}^{film} (Bi_{1-x} 
Sb_{x} )} \cdot E_{g}^{bulk} (Bi) 
\end{equation} 
and
\begin{equation} \label{GrindEQ__6_} 
E_{g}^{film} (Bi_{1-x} Sb_{x} )=E_{g}^{bulk} (Bi_{1-x} Sb_{x} )+\frac{\pi^{2} \alpha_{33}^{film} (Bi_{1-x} Sb_{x} )}{l_{z}^{2} } .                
\end{equation} 
The dispersion relation $E(\mathbf{k})$ can then be the solved by the methods used by Ref. \cite{12} from
\begin{equation} \label{GrindEQ__7_} 
E(\mathbf{k})+\frac{E^{2} (\mathbf{k})}{E_{g}^{film} } =\frac{1}{2}{\mathbf{k}}^{*} \cdot \tilde{\bm\upalpha }\cdot {\mathbf{k}}+
\frac{\pi^{2} \alpha _{33} }{2 l_{z}^{2} } ,                
\end{equation} 
where 
$\tilde{\alpha }_{ij} =\alpha _{i3}\alpha _{j3}/\alpha_{33} -\alpha _{ij}$ for $i,j=1$ and $2$, and $\bm\upalpha =\bm\upalpha ^{film} (Bi_{1-x}Sb_{x} )$. The Hamiltonian for Bi and Bi${}_{1-x}$Sb${}_{x}$ based on $\mathbf{k}\cdot \mathbf{p}$ theory in Eq. \eqref{GrindEQ__2_} is equivalent to a Dirac Hamiltonian 
with a scaled canonical conjugate momentum \cite{14}. Thus, Eq. \eqref{GrindEQ__7_} is also a good approximation to describe the Dirac cones. The band parameters we use in the present calculations are values that were measured by cyclotron resonance experiments \cite{15}.

\indent According to Eqs. \eqref{GrindEQ__1_} and \eqref{GrindEQ__7_}, when $E_{g} \to 0$ at an \textit{L} point, the electronic dispersion relation becomes a perfect Dirac cone, where the energy $E$ is exactly proportional to the lattice momentum $\mathbf{k}$ measured from that \textit{L} point. When $E_{g}$ becomes large enough \cite{f1}, the linearity of the dispersion relation becomes an approximation, and the Dirac cone becomes a quasi-Dirac cone. If $\tilde{\alpha }_{11} \gg \tilde{\alpha }_{22} $ with a finite $E_{g} $, so that $E\propto k_{x} $ and $E\propto k_{y}^{2} $, we call it a semi-Dirac cone. In a semi-Dirac cone, the fermions are relativistically dispersed in one direction ($k_{x}$), and classically dispersed in another direction ($k_{y}$).
 
\begin{figure}
        \includegraphics[width=0.5\textwidth]{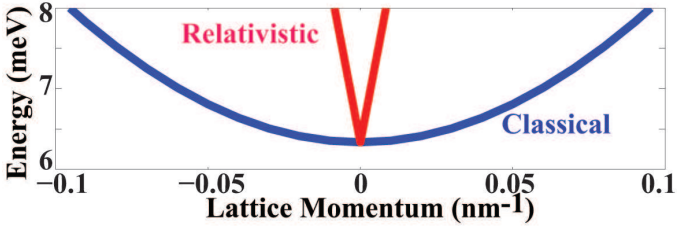}
        \caption{Example of a semi-Dirac cone in the Bi${}_{1-x}$Sb${}_{x}$ thin film system ($x=0.10$ and $l_{z} =100$ nm). It can be seen that around the $L^{(1)}$ point, the fermions are relativistic (linearly dispersed) along the  $\mathbf{v}_{\max } $ direction, and classical (parabolically dispersed) along the $\mathbf{v}_{\min } $ direction.}
\end{figure}

\indent We propose that single-, bi- and tri-Dirac-cone materials can be constructed from Bi${}_{1-x}$Sb${}_{x}$ thin films, by proper synthesis conditions to control the relative symmetries of the 3 \textit{L} points.  Bi${}_{1-x}$Sb${}_{x}$ thin films grown along the bisectrix axis can be single-Dirac-cone materials, as illustrated in Fig. 2a, where the 3-fold degeneracy of the $L^{(1)} $, $L^{(2)} $ and $L^{(3)} $ points is broken. The value of the film-direction-inverse-mass-component $\alpha _{33}^{film} (Bi_{1-x} Sb_{x} )$ is much smaller for the $L^{(1)}$ point 
than the corresponding values for the $L^{(2)} $ and $L^{(3)} $ points. The $L^{(1)} $-point gap $E_{g}^{(1)} $ is negligibly small
due to the small value of $\alpha _{33}^{film} (Bi_{1-x} Sb_{x} )$, where a Dirac cone is formed, as shown in Fig. 2a. However, the $L^{(2)} $- and $L^{(3)} $- point band gaps $E_{g}^{(2)} $ and $E_{g}^{(3)} $ are much 
larger, which implies that a single-Dirac-cone at the $L^{(1)} $ point is constructed. Here we are taking advantage of both 
the extreme anisotropy of Bi${}_{1-x}$Sb${}_{x}$ and the quantum confinement effect of thin films. The quantum confinement effects for the $L^{(1)} $-point carriers differs remarkably from those for the $L^{(2)} $- and $L^{(3)} $-point carriers due to the anisotropy of the \textit{L}-point pockets. Figure 2b shows that a Bi${}_{1-x}$Sb${}_{x}$ thin film grown along the binary axis can be a bi-Dirac-cone material, where the $L^{(1)} $-point band gap $E_{g}^{(1)}$ is much larger than the $L^{(2)} $- and $L^{(3)} $- point band gaps $E_{g}^{(2)} $ and $E_{g}^{(3)}$. Thus, $E_{g}^{(2)}$ and $E_{g}^{(3)} $ remain small enough \cite{f1} to make two degenerate Dirac cones (quasi-Dirac cones) at the $L^{(2)} $ and 
the $L^{(3)} $ points. In Fig. 2c, the film is grown along the trigonal axis, so that the 3-fold 
symmetry for the three \textit{L} points is retained. The three Dirac cones (quasi-Dirac cones) at the $L^{(1)} $, $L^{(2)} $ and $L^{(3)} $ points are degenerate in energy, which makes this film a tri-Dirac-cone material. By definition, an exact Dirac cone has $E_g=0$. However, $E_g=0$ Dirac cones are seldom achieved experimentally, so it is practical to consider $E_g \le k_B \do T$ as a criterion for an exact Dirac cone. In the temperature range below 77 K that we are considering in this paper, the thermal smearing of $k_B \do T$ corresponds to $\sim 7$ meV. For the criterion of a quasi-Dirac cone, we can use $k_B \do T \le E_g \le E_g(Bi)^{bulk}$, where $E_g(Bi)^{bulk} \simeq 14$ meV. Thus, we consider the three Dirac cones in Fig. 2c, as quasi-Dirac cones, which are plotted for the case of $l_z=100$ nm and  $E_g\simeq 10$ meV. If exact Dirac cones are needed, a larger film thickness can be chosen, e.g. $l_z=200$ nm, which satisfies $E_g \le k_B \do T$. 
 
\indent We now show how to construct anisotropic Dirac cones with different shapes for the wave vector as a function of cone angle. To characterize the anisotropy 
of a Dirac cone, we define an anisotropy coefficient 
\begin{equation} \label{GrindEQ__8_} 
\gamma =\frac{|\mathbf{v}_{\max}| }{|\mathbf{v}_{\min }|} ,                
\end{equation} 
where $\mathbf{v}_{\max } $ and $\mathbf{v}_{\min } $ are the maximum and minimum in-film carrier group velocities 
for a Dirac cone that is defined as $\mathbf{v}(\mathbf{k})= \nabla _{\mathbf{k}} E(\mathbf{k})$. For a perfect Dirac cone, $\mathbf{v}$ is a function of the direction of the lattice momentum $\mathbf{k}$ measured 
from that \textit{L} point only and is independent of the magnitude of $\mathbf{k}$. For an imperfect Dirac cone or a quasi-Dirac cone, this magnitude invariance is exact only when $\mathbf{k}$ is large, and becomes an approximation around the apex when $\mathbf{k}$ is small \cite{f1}. 

\indent Fig. 3 gives us an important guide on how to construct anisotropic $L^{(1)}$-point Dirac cones. In Fig. 3a, the anisotropy coefficient $\gamma $ for the $L^{(1)} $-point Dirac cone as a function of film growth orientation is shown. For a film grown along the bisectrix axis, $\gamma $ has its minimum value $\gamma_{min}= \tiny{\sim}{2}$, where the carrier velocity $\mathbf{v}(\mathbf{k})$ for the $L^{(1)}$-point Dirac cone varies only by a small amount with the direction of $\mathbf{k}$, as shown in Fig. 3b. For a film grown along the binary axis, $\gamma=\tiny{\sim}{10}$, where $\mathbf{v}(\mathbf{k})$ varies more with the direction of $\mathbf{k}$ as shown in Fig. 3c, compared to Fig. 3b. For a film grown along the trigonal axis, $\gamma$ has its maximum of $\gamma_{max}=\tiny{\sim}{14}$, where $\mathbf{v}$ varies significantly with the direction of $\mathbf{k}$, as shown in Fig. 3d. 

\indent Researchers have tried to realize semi-Dirac cones in oxide layers \cite{16}, where the fermions are relativistic in one direction and classical in its orthogonal direction. In the present work, we have found that it is possible to construct semi-Dirac cones in the Bi${}_{1-x}$Sb${}_{x}$ thin film system.  According to Eqs. \eqref{GrindEQ__1_}, \eqref{GrindEQ__2_} and \eqref{GrindEQ__7_}, for an in-film direction $\hat{\mathbf{k}}$, where $\hat{\mathbf{k}}$ is a unit directional vector of $\mathbf{k}$ in the in-film lattice momentum space,  whether the dispersion relation is linear or parabolic depends on the \textit{L}-point band gap $E_{g} $, and the $\tilde{\bm\upalpha}$ projection along that direction of $\hat{\mathbf{k}}$, defined by $\tilde{\alpha }_{\hat{\mathbf{k}}}=\hat{\mathbf{k}}^{*} \cdot \tilde{\bm\upalpha }\cdot \hat{\mathbf{k}}$, where $\tilde{\bm\upalpha}$ is given by Eq. \eqref{GrindEQ__7_}.  When $E_{g} $ is small and $\tilde{\alpha }_{\hat{\mathbf{k}}} $ is large \cite{f1}, the energy becomes linearly dispersed along $\hat{\mathbf{k}}$; when $E_{g} $ is large and $\tilde{\alpha }_{\hat{\mathbf{k}}} $ is small, the energy becomes parabolically dispersed 
along $\hat{\mathbf{k}}$. To construct a semi-Dirac cone, we needs to find a proper \textit{L}-point band gap $E_{g} $ and anisotropy $\gamma $,  such that $E_{g} /\tilde{\alpha }_{
\max } $ is small and $E_{g} /\alpha _{\min } $ is large. In this case, the electronic energy is linearly dispersed along the $\tilde{\alpha }_{\max } $ direction and parabolically dispersed along the $\alpha _{\min } $ direction. Here $\tilde{\alpha }_{\max } $ and $\tilde{\alpha }_{\min } $  are the maximum and minimum values of $\tilde{\alpha }_{\hat{\mathbf{k}}} $, which correspond to the principal axes of the 2D tensor $\tilde{\bm\upalpha }$.  The \textit{L}-point 
band gap varies as a function of the film thickness \textit{l}${}_{z}$, the growth orientation and the Sb composition \textit{x}, as shown by the calculated results given in Fig. 4. To construct 
a semi-Dirac cone, we need to find a growth direction that ensures a significant anisotropy, 
and a large enough value of $E_{g} $ which ensures that $E(\mathbf{k})$ becomes parabolically dispersed along the $\tilde{\alpha }_{\min} $ direction. However, the $E_{g} $ should not be 
too large, because of the necessity that the linear dispersion relation along the $\tilde{\alpha }_{\max } $ direction is maintained. These requirements can all be achieved by choosing the proper Sb composition \textit{x}, film thickness \textit{l}${}_{z}$ and growth orientation as shown in Fig. 4. From Figs. 2 and 3, we know that the $L^{(1)} $-point Dirac cone has a maximum \textbf{k}-vector anisotropy, when the growth orientation is near the trigonal axis. We also see that the thin film becomes a single-Dirac-cone material when the growth direction is near the bisectrix axis. Thus, a good strategy to construct a semi-Dirac cone is to choose a growth orientation between the trigonal and the bisectrix axis, in the trigonal-bisectrix plane. An example of a semi-Dirac cone is shown in Fig. 5, where the example sample is grown along a direction that is $40^{\circ } $ from the trigonal axis, $50^{\circ}$ from the bisectrix axis, and perpendicular to the binary axis. Thus, a large Sb composition (e.g. $x\approx 0.10$) and a small film thickness (e.g. $l_z\approx 100$ nm) is preferred to make the $E_g$ large, and $x=0.10$ and $l_{z} =100$ nm are chosen for this example sample. 

\indent In conclusion, we have proposed the growth of Bi${}_{1-x}$Sb${}_{x}$ thin films, which for selected concentrations of Sb and different directions to the film normal allows different Dirac-cone materials to be constructed. We have shown how to construct single-, bi- and tri-Dirac-cone materials, as shown in  Fig. 2, as well as quasi- and semi-Dirac-cone materials, as shown in Fig. 2c and Fig. 5, respectively.

\begin{acknowledgments}
The authors acknowledge the support from AFOSR MURI Grant number FA9550-10-1-0533, sub-award 60028687. The views expressed are not endorsed by the sponsor.
\end{acknowledgments}


\begin{thebibliography}{10000}
\bibitem{f1}   
Based on Eq. (1), when $E_g$ is large enough of $\mathbf{k}$ is small enough so that $E_g \ll |\mathbf{v} \cdot \mathbf{k} |$, $E(\mathbf{k})$ can be considered as linearly dispersed along $\mathbf{k}$.
\end{thebibliography}
\end{document}